\documentclass[prd,preprintnumbers,floatfix,aps,nofootinbib,notitlepage]{revtex4}

\usepackage{latexsym}
\usepackage{epsfig}
\usepackage{amssymb}

\newcommand{\lp}{\left(}
\newcommand{\rp}{\right)}
\newcommand{\lb}{\left[}
\newcommand{\rb}{\right]}

\newcommand{\ba}{\begin{eqnarray}}
\newcommand{\ea}{\end{eqnarray}}
\newcommand{\be}{\begin{equation}}
\newcommand{\ee}{\end{equation}}

\newcommand{\al}{\alpha}
\newcommand{\bt}{\beta}
\newcommand{\ga}{\gamma}
\newcommand{\ka}{\kappa}

\newcommand{\la}{\lambda}

\newcommand{\C}{\mathcal{C}}
\newcommand{\R}{\mathcal{R}}

\newcommand{\beq}{\begin{equation}}
\newcommand{\eeq}{\end{equation}}
\newcommand{\bea}{\begin{eqnarray}}
\newcommand{\eea}{\end{eqnarray}}

\begin{document}

\title{Non-metric chaotic inflation}

\author{Kari Enqvist}
\email{kari.enqvist@helsinki.fi}
\affiliation{Physics Department, University of Helsinki, and Helsinki Institute of Physics, FIN-00014 University of Helsinki.}
\author{Tomi Koivisto}
\email{T.S.Koivisto@uu.nl}
\affiliation{Institute for Theoretical Physics and Spinoza Institute, Leuvenlaan 4, 3584 CE Utrecht, The Netherlands.}
\author{Gerasimos Rigopoulos}
\email{rigopoulos@physik.rwth-aachen.de}
\affiliation{\it Institut f\"ur Theoretische Teilchenphysik und Kosmologie,\\
RWTH Aachen University,\\
D--52056 Aachen, Germany}

\date{\today}

\begin{flushright}
{
\small
HIP-2011-20/TH\\ ITP-UU-11/28 \\ SPIN-11/21\\TTK-11-25
}
\end{flushright}

\begin{abstract}
We consider inflation within the context of what is arguably the simplest non-metric extension of Einstein gravity.
There non-metricity is described by a single graviscalar field with a non-minimal kinetic coupling to the inflaton field  $\Psi$, parameterized by a single parameter $\gamma$.
We discuss the implications of non-metricity for chaotic inflation and find that it significantly alters the inflaton dynamics for field values $\Psi \gtrsim M_P/\gamma$, dramatically changing the qualitative behaviour in this regime.
For potentials with a positive slope non-metricity imposes an upper bound on the possible number of e-folds.
For chaotic inflation with a monomial potential, the spectral index and the tensor-to-scalar ratio receive small corrections dependent on the non-metricity parameter. We also argue that significant post-inflationary non-metricity may be generated.

\end{abstract}

\maketitle

\section{Introduction}

Even at the classical level, the theory of gravity has still two major open issues. First, what is the form of the action?
Here investigations have focused on extensions of Einstein gravity, such as $f(R)$ gravities \cite{Sotiriou:2008rp}, and their implications
for dark energy or inflation \cite{Nojiri:2006ri,Clifton:2011jh}. But there is also the second, and arguably even more profound question: what are the true gravitational degrees of freedom? Conventionally, one just assumes
metricity: that gravity can be described by the metric alone, setting the connection $\hat\Gamma$ to be the Levi-Civita connection with $\hat\Gamma=\Gamma[g_{\mu\nu}]$ by hand.
In the Palatini variation one takes the metric and the connection to be independent degrees of freedom \cite{Hehl:1994ue}, and varying both
results in equations of motion that differ from the metric case except if the Lagrangian has exactly the Einstein-Hilbert
form with $f(R)=R$. In the general Palatini case, the solution to equation of motion for the connection is not the Levi-Civita
connection, and hence such theory is called non-metric. In non-metric gravity, the metric is no longer covariantly
conserved.

However, several issues arise in the Palatini approach \cite{Olmo:2011uz}, which may stem from the nontensorial property of the connection.
Instead, one could consider
theories with two metrics: one for describing the geometry of the manifold, and the other for generating the connection.
Indeed, the Palatini case can be seen as a special case of a more general class of conformal non-metric theories.
In these C-theories \cite{Amendola:2010bk,Koivisto:2011vq} one postulates a metric for the connection, $\hat g_{\mu\nu}$
that is related to the gravitational metric by a conformal factor $\C$ with $\hat g_{\mu\nu}=\C g_{\mu\nu}$, the Palatini-f(R) then
corresponding to the choice $\C(R)=f'(R)$.
The connection is then the Levi-Civita connection of the connection-generating metric with $\hat\Gamma=\Gamma[\hat g_{\mu\nu}]$.
Such a relation was
originally considered by Weyl in his conformally invariant theory of gravity \cite{0264-9381-13-1-013}.

Surprisingly, the subtle refinement of the fundamental spacetime structure turns out to have concrete and
interesting consequences, for one allowing non-metric degrees of freedom to propagate in vacuum. This is in contrast both to the result of the standard metric variation (where the connection is fixed to be metric a priori) and the metric-affine, i.e. Palatini, variation (where the connection is constrained a posteriori), which however can be recovered as specific limits of C-theories. Thus, they provide a unified approach
to a variety of previously seemingly very different alternative gravity theories but reveal also novel possibilities hidden already in the Einstein-Hilbert action itself.


The nature of gravity is of great relevance for models of inflation \cite{Mazumdar:2010sa}. 
Since inflation occurs at very high curvature scales where one naturally expects corrections to Einstein's theory, the ambiguities both in the
form of gravitational Lagrangian and its degrees of freedom should be taken into account.
The purpose of this paper is to explore the implications of the simplest conformal non-metricity to inflationary physics. In particular, we analyze the corrections that inevitably appear to large-field chaotic inflation.
Our approach is different from considering inflation to be driven by non-metricity,
which has been pursued in Refs. \cite{0264-9381-5-3-013,0264-9381-8-5-014}. The Palatini variation has been applied to a non-minimally coupled inflation \cite{Bauer:2010jg,Tamanini:2010uq}, however there taking only into account the shift of the kinetic term \cite{Koivisto:2005yc}, whereas our purpose is to consider the generic lagrangian (for details, see appendix \ref{quadratic}).

In the next section, we couple a matter scalar field into gravity in the presence of Weyl non-metricity and derive the equations of motion for homogeneous fields. In the following section we then analyze the dynamics of slow-roll chaotic inflation and derive constraints on the amount of non-metricity chaotic inflation can tolerate. The results are discussed in section \ref{conclusions}. The generic starting lagrangian and the general field equations are confined to the appendices.

\section{Scalar field and non-metric gravity}

Let us now consider the possibility that the spacetime connection $\hat{\Gamma}$ is conformally related to the
Levi-Civita connection $\Gamma$ with
\be \label{conformal}
\hat{\Gamma}^\alpha_{\bt\ga} = {\Gamma}^\alpha_{\bt\ga} + \lp
\delta^\al_{(\bt}\delta^\lambda_{\ga)}-\frac{1}{2}g_{\bt\ga}g^{\al\lambda}\rp\nabla_\lambda \log\C\,,
\ee
where $\C=\C(x)$ is the conformal factor.
The scalar curvature is then\footnote{Here $\Box$ is the metric d'Alembertian and other unhatted quantities are
also metric unless otherwise specified. The reduced Planck mass is denoted by $\kappa \equiv \sqrt{8\pi G}$.}
\be \label{ricci}
\R \equiv g^{\mu\nu}\hat{R}_{\mu\nu} = R - \frac{D-1}{4\C^2}\lb 4\C \Box\C+(D-6)(\partial \C)^2\rb\,.
\ee
The conformal factor $\C$ represents a scalar degree of freedom. If we define $\varphi$ by
\be
\C \equiv e^{\frac{2\kappa}{\sqrt{(D-1)(D-2)}}\varphi}\,,
\ee
the Einstein-Hilbert action can be written simply as
\be
S_{E-H}=\frac{1}{2\kappa^2}\int d^Dx \sqrt{-g}\R = \int d^Dx \sqrt{-g}\lb \frac{R}{2\kappa^2} -(\partial\varphi)^2 \rb\,,
\ee
so the conformal relation contributes, effectively, a massless scalar field to the vacuum. This is a remarkably
simple way of introducing non-metricity and allows to study its effect on the coupling of gravity and matter.

For this purpose, let us consider a scalar matter field $\Psi$ with a mass $m$.
With the canonical action for the matter field, the theory is metric, though more general than Einstein's theory \cite{Koivisto:2005yk}.
In the presence of non-metricity, one has to reconsider the coupling of the scalar matter fields to gravity.
There are several possible lagrangians that all
reduce to a canonical scalar field theory when the non-metricity vanishes, i.e. when $\C$ is a constant.
The corresponding action can then be interpreted as the work done along the path of the scalar particle, but now
the free falling trajectories are affected by the non-metricity of the connection. Hence the matter fields are
coupled not only to the metric but also to the connection, or equivalently, to the graviscalar field $\varphi$.

In the appendix \ref{quadratic}
it is shown that the new effects in our case are captured by just a single term, with an unknown coefficient $\gamma$ in the effective lagrangian which becomes
\be \label{theory}
2L = \frac{1}{\kappa^2} R - \lp\partial\varphi\rp^2
- \lp\partial \Psi\rp^2 + 2\gamma \kappa\Psi\Psi_{,\al}\varphi^{,\al} - 
V(\Psi)~,
\ee
where $\kappa=1/M_P$. We thus obtain effectively a two-field system with field dependent kinetic term. The non-metric effects are suppressed by the Planck scale $M_P$.
The most general lagrangian is considered in the Jordan and in the Einstein frame in the appendix \ref{quadratic}.

Let us now study the implications of the non-metric graviscalar $\varphi$ for chaotic inflation. The equations of motion for the
homogenous scalar $\Psi$ and graviscalar $\varphi$ fields in FRW universe are given by
\ba
\ddot{\varphi}+3H\dot{\varphi} -  \frac{\ga\kappa}{1-(\ga\ka\Psi)^2}\dot{\Psi}^2  + 
\frac{2\ga\kappa}{1-(\ga\ka\Psi)^2}\,V & = & 0\,, \label{varphi1}\\
\ddot{\Psi} + 3H\dot{\Psi} -  \frac{\ga^2\kappa^2}{1-(\ga\ka\Psi)^2}\Psi\dot{\Psi}^2  +  
\frac{V_{,\Psi}}{1-(\ga\ka\Psi)^2} & = & 0\,, \label{psi1}
\ea
where the $\dot{\Psi}^2$ terms arise from the field dependent metric in the kinetic term of (\ref{theory}). We can bring the field metric into a flat form by defining new fields $u_\pm$ and $v$ through
$du_-=\sqrt{1-\gamma^2\kappa^2\Psi^2}d\Psi$ (for the region $\Psi<1/\gamma\kappa$), $du_+=\sqrt{\gamma^2\kappa^2\Psi^2-1}\,d\Psi$
(for the region $\Psi>1/\gamma\kappa$),
and $dv=d\varphi-\gamma\kappa\Psi \, d\Psi$.
Then the scalar degrees of freedom are given by
\ba
u_-&=&\Psi\frac{\sqrt{1-\gamma^2\kappa^2\Psi^2}}{2}+\frac{\arcsin(\gamma\kappa\Psi)}{2\gamma\kappa}\,\quad (\Psi<1/\gamma\kappa),\\
u_+&=&\frac{\pi}{4\gamma\kappa}
+\Psi\frac{\sqrt{\gamma^2\kappa^2\Psi^2-1}}{2}-\frac{\ln\left(\gamma\kappa\Psi+\sqrt{\gamma^2\kappa^2\Psi^2-1}\right)}{2\gamma\kappa}
\,\quad (\Psi>1/\gamma\kappa),\\
v&=&\varphi-\frac{1}{2}\gamma\kappa\Psi^2~.
\ea
In terms of the redefined fields the equations of motion now read
\bea\label{u1}
&&\ddot{u}_{\mp}+3H\dot{u}_{\mp}\pm\frac{V_{,\Psi}(u_{\mp})}{\sqrt{|1-\gamma^2\kappa^2\Psi^2(u_{\mp})|}}
 = 0\\
&&\ddot{v}+3H\dot{v}=0\,,\\
&& H^2= \frac{1}{3}\left[\frac{1}{2}\dot{v}^2 \pm \frac{1}{2}\dot{u}_{\mp}^2+
V(u_{\mp})\right]\,.
\eea
We immediately see that non-metricity qualitatively alters the field dynamics when $\gamma\kappa u\gtrsim\pi/4$ ($\Psi\gtrsim 1/\gamma\kappa$). In particular, for $\gamma\kappa u\gtrsim\pi/4$ (the $+$ region) an attractive potential $V$ translates into a repulsive force on the field and vice versa. Taking for concreteness a monomial potential $V(\Psi)=\lambda\kappa^{\alpha-4}\Psi^\alpha$ we see that the Klein-Gordon equation (\ref{u1}) becomes
\beq\label{KG}
\ddot{u}_{\mp}+3H\dot{u}_{\mp}\pm\frac{\alpha\lambda}{\kappa^3}\frac{1}{\gamma^{\alpha-1}}
\frac{\left(\gamma\kappa\Psi\right)^{\alpha-1}}{\sqrt{|1-\gamma^2\kappa^2\Psi^2(u_{\mp})|}}
= 0
\eeq
where close to the point $\gamma\kappa u=\pi/4$ we have
\beq\label{approx}
\gamma\kappa\Psi \simeq 1+{\rm Sign}\left(\gamma\kappa u-\frac{\pi}{4}\right)\left(\frac{3}{2\sqrt{2}}\right)^{2/3}\left|\gamma\kappa u-\frac{\pi}{4}\right|^{2/3}
\eeq

For potentials $V$ with a positive slope we see that a barrier forms at this point, separating the $-$ from the $+$ region. The force exerted there is infinite and the field is driven away from $ \Psi =1/\gamma\kappa$. In the $+$ region the field then grows without bound and is also a ghost. Even though solutions exist as long as its kinetic term doesn't dominate, it is difficult to see how this would lead to a sensible cosmology. On the other hand, the $-$ region does lead to sensible slow roll inflation. Slow roll breaks down close to the barrier and hence the total number of efolds in such a model is bounded. Note that even though the force is infinite, the barrier has finite width and height and is thus in principle crossable both classically and quantum mechanically. If the potential has a negative slope in the $+$ region the force is attractive (but again infinite at $ \Psi =1/\gamma\kappa$). One can also see that the solution $v={\rm const}$ is an attractor. This fixes the graviscalar in terms of the inflaton to be
\beq
\varphi=\frac{1}{2}\gamma\kappa\Psi^2\,.
\eeq

\begin{figure}[t]
\includegraphics[scale=0.7]{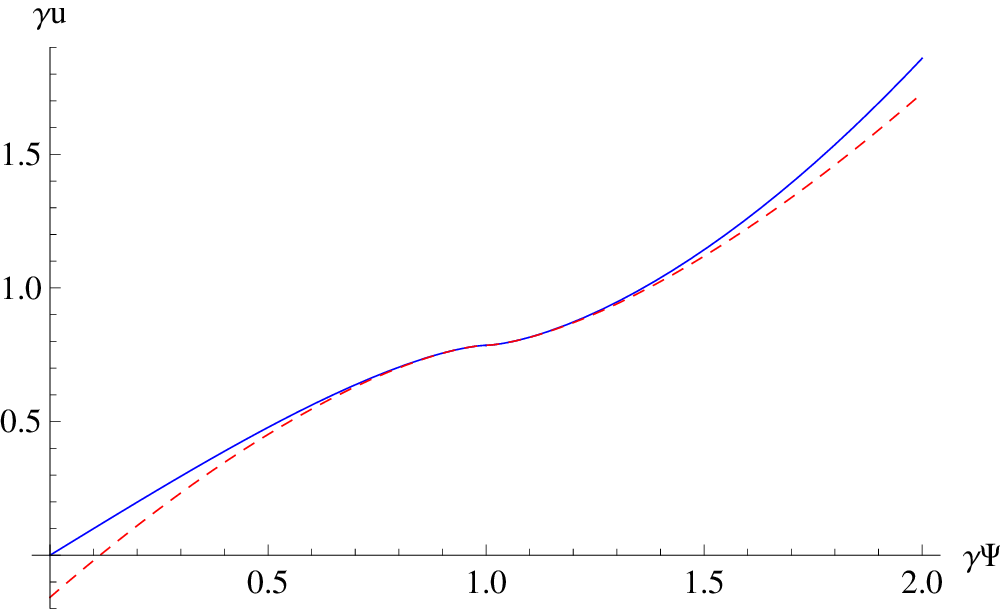}\hspace{1cm}
\includegraphics[scale=0.7]{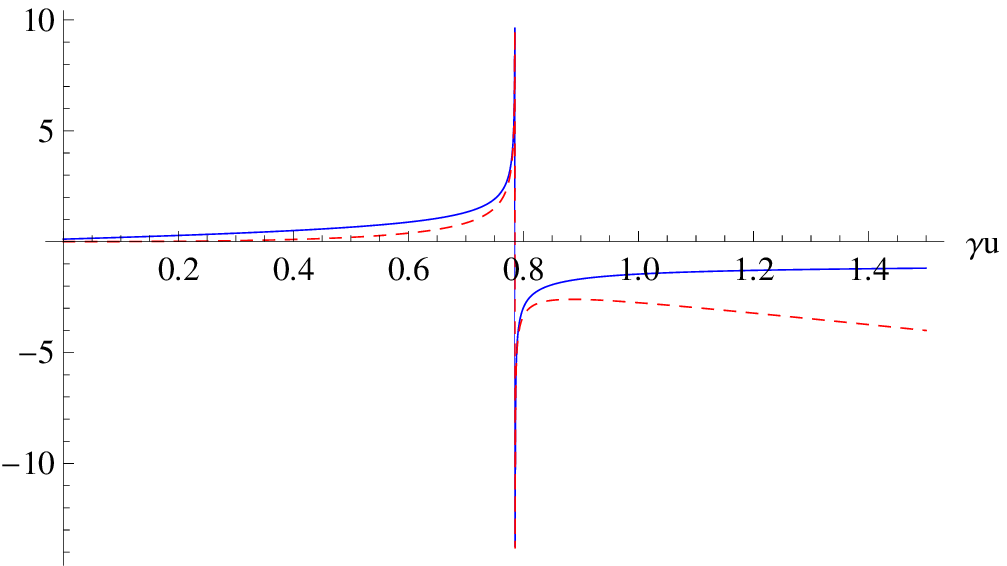}
\label{spherical}
\caption{Left: The field $u$ is plotted as a function of $\Psi$. Inversion of this graph gives $\Psi(u)$. The dashed line shows the approximation (\ref{approx}). Right: The third term of the field equation (\ref{KG}) (the negative of the ``force'' felt by the $u$ field) for $\alpha=2$ (solid) and $\alpha=4$ (dashed). A repulsive barrier is formed at the point $\gamma u=\pi/4$. ($M_{\rm P}$ is set to unity in this figure.)}
\end{figure}


\section{Slow-roll chaotic inflation}

Let us now consider slow-roll chaotic inflation with a monomial potential $V(\Psi)=\lambda\kappa^{\alpha-4}\Psi^\alpha$. The slow roll equations of motion then read
\beq
3H\dot{u}\simeq-\frac{\lambda\alpha\kappa^{\alpha-4}\Psi(u)^{\alpha-1}}{\sqrt{1-\gamma^2\kappa^2\Psi(u)^2}}\,,\quad H^2\simeq\frac{\kappa^2}{3}\lambda\kappa^{\alpha-4}\Psi(u)^\alpha~
\eeq
while $v={\rm const}$. The inflaton, $u$ starts to roll with some initial value $u < \pi/8\gamma\kappa$. Using $N$ as the time parameter, and reverting from $u$ back to $\Psi$ for the moment, we obtain for the number of e-folds
\beq
\frac{d\Psi}{d N} \simeq \frac{\alpha}{\kappa^2}\frac{1}{\Psi\left(1-\gamma^2\kappa^2\Psi^2\right)}\,,
\eeq
which is easily integrated to give
\beq
N\simeq\frac{\kappa^2}{\alpha}\frac{\Psi^2}{2}\left(1-\frac{\gamma^2\kappa^2}{2}\Psi^2\right)\,.
\eeq
The approximate equality holds after ignoring a constant of integration and focusing on field values relevant for cosmological observations. Slow roll inflation can be supported in the region $\dot{u}^2<2V(\Psi)$, which implies that slow roll breaks down at field values
\beq\label{signs}
\Psi^2=\frac{1\mp\sqrt{1-\frac{2\alpha^2}{3}\gamma^2}}{2\gamma^2\kappa^2}\,.
\eeq
We see that the necessary condition for the slow roll regime to exist at all is $\gamma < \sqrt{{3}/{2\alpha^2}}$ and that slow roll
breaks down for two field values: the minus sign in (\ref{signs}) gives a lower bound on the value of $\Psi$, which in the limit $\gamma \rightarrow 0$ reproduces the standard result $\Psi_1^2\simeq\alpha^2/6\kappa^2$. However, there is also an upper bound for the field value beyond which slow roll is no longer possible. This is because when $\Psi\sim 1/\gamma\kappa$, non-metricity introduces corrections which make the effective potential felt by the inflaton field steeper, rendering slow roll unattainable. Hence, we conclude that non-metricity introduces a bound to the possible number of e-folds a chaotic inflation monomial potential can give. A similar result holds for any slow roll potential with a positive slope. Assuming $\gamma\ll 1$, we find that for the monomial potential
\beq\label{efolds1}
N<\frac{1}{4\alpha\gamma^2}\,,
\eeq
while requiring that $N \gtrsim 60$ restricts $\gamma$ to be
\beq
\gamma \lesssim \frac{1}{16\sqrt{\alpha}}\,.
\eeq

Deep in the slow roll region where $u \ll\pi/8\gamma\kappa $ we can approximate
\beq
\Psi(u)\simeq u+\frac{1}{6}\gamma^2\kappa^2u^3
\eeq
whence the equation of motion reads
\beq
\ddot{u}+3H\dot{u} + \alpha\lambda\kappa^{\alpha-4}u^{\alpha-1}\left(1+\frac{2+\alpha}{6}\gamma^2\kappa^2u^2\right) =0.
\eeq
This corresponds to an effective potential
\beq\label{effpotforinf}
V(u)=\lambda\kappa^{\alpha-4}u^\alpha\left(1+\frac{\alpha}{6}\gamma^2\kappa^2u^2\right)
\eeq
and
\beq
H^2=\frac{\kappa^2}{3}\left[\frac{\dot{v}^2}{2}+\frac{\dot{u}^2}{2}
+V(u)\right]
\eeq
Thus, in the slow-roll region non-metricity appears as a subdominant correction in the effective potential felt by the inflaton scalar. The number of e-folds as a function of $U$ is calculated to be
\beq
N\simeq\frac{\kappa^2}{2\alpha}u^2\left(1-\frac{1}{6}\gamma^2\kappa^2u^2\right)
\eeq
from which $u$ can be written as a function of $N$
\beq
\kappa^2u^2\simeq 2\alpha N \left(1+\frac{\alpha}{3}\gamma^2N\right)
\eeq
From the above results we obtain for the spectral index at order $\gamma^2$
\beq\label{spectral-index}
n-1=-6\epsilon+2\eta=-\frac{2+\alpha}{2N}-\frac{\alpha(\alpha-2)}{2}\gamma^2\,.
\eeq
Interestingly, the correction at this order vanishes for $\alpha=2$. The tensor to scalar ratio is calculated to be
\beq\label{tensor-to-scalar}
r\simeq 12.4\epsilon\simeq 3.1 \alpha\left(\frac{1}{N}+\alpha\gamma^2\right)
\eeq

Let us immediately note that the calculations for the spectral index and the tensor to scalar ratio, equations (\ref{spectral-index}) and (\ref{tensor-to-scalar}), ignore the isocurvature perturbation associated with the extra degree of freedom. Although a proper treatment requires a linear perturbation analysis starting from the initial lagrangian (\ref{theory}), we can make the following estimates: In the $\{u,v\}$ basis inflation proceeds entirely along the $u$ direction, which definines the adiabatic perturbation, while the isocurvature field corresponds to $v$. Since the trajectory is straight there is no coupling between the two at the linear level. Thus we expect the single field conclusions (\ref{spectral-index}) and (\ref{tensor-to-scalar}) to hold for the adiabatic perturbation related to $u$, or equivalently $\Psi$, while the perturbations $\delta v(k) \sim H/k^{3/2}$ would imply a perturbation for the graviscalar
\beq\label{isocurv}
\delta\varphi(k) \simeq \left(1+\gamma\kappa\Psi_*\right)\frac{H_*}{k^{3/2}}
\eeq
where the subscript $*$ as usual indicates evaluation at horizon exit. We discuss the implications of this briefly in the discussion.

\section{Discussion}
\label{conclusions}
The simplest possible implementation of a non-metric theory of gravity introduces a new scalar degree of freedom associated with the conformal part of the metric. In turn, this can be cast as a theory of two scalar fields with a non-trivial metric in field space. We have examined the implications of such a non-metric theory for inflation and found that for inflaton values $\Psi\gtrsim 1/\gamma\kappa$ such a seemingly innocuous modification has profound implications for the way inflation proceeds. In particular, as can be seen from eq (\ref{u1}) the effect of non-metricity is to effectively turn an attractive inflaton potential into a repulsive one and viceversa. Thus, an inflationary potential with a positive slope becomes unstable in the region $\Psi\gtrsim 1/\gamma\kappa$ leading to a cosmology that cannot represent the inflationary past of our universe. On the contrary, for values $\Psi < 1/\gamma\kappa$ slow roll inflation can proceed with few modifications, mainly in the form of small $\gamma$ corrections for the spectral index and the scalar to tensor ratio. The main difference from metric inflation is that now the non-metricity parameter $\gamma$ sets un upper bound for the possible number of e-folds, see eq (\ref{efolds1}). In this sense non-metricity can act as an IR regulator for inflation.

It is perhaps interesting to note that potentials which are repulsive can turn attractive in the $\Psi\gtrsim 1/\gamma\kappa$ region which could then support a viable inflationary cosmology, although again with a point of infinite force for the field at $\Psi= 1/\gamma\kappa$.

Let us close by briefly discussing the effects of perturbations. We argued that non-metric inflation would give rise to a fluctuation of the graviscalar $\varphi$ in addition to the standard adiabatic metric perturbation, see eq (\ref{isocurv}). After the inflaton decays the graviscalar perturbation would presumably persist along with the adiabatic metric one $\zeta(k)\simeq{H_*}/{\sqrt{\epsilon}k^{3/2}}$. This would give rise to a non-metric contribution to the perturbed part of the connection, see equation (\ref{conformal}), with $(\hat{\Gamma}-\Gamma)/\Gamma\sim\sqrt{\epsilon}$. Thus, non-metric inflation seems to be generating deviations of the perturbed post-inflationary connection from the metric case. Such deviations could be used to constrain non-metricity further and we will return to this issue in future work.

\subsubsection*{Acknowledgements}

\noindent

KE is supported by the Academy of Finland grants 218322 and 131454.
GR is supported by the Gottfried Wilhelm Leibniz programme
of the Deutsche Forschungsgemeinschaft.

\appendix

\section{General scalar field couplings}
\label{quadratic}

The most general lagrangian for a scalar field $\Psi$ that is parity-invariant, first order in curvature and at most quadratic in
derivatives of $\Psi$ can be specified by six functions $A_i(\Psi)$, $i=1,\dots, 6$ as:
\be \label{lagrangian}
L=A_1(\Psi)\R + A_2(\Psi)(\partial\Psi)^2 + A_3(\Psi)\Psi^{,\nu}g_{\al\bt}\hat{\nabla}_\nu g^{\al\bt}
+A_4(\Psi)\Psi_{,\mu}\hat{\nabla}_\nu g^{\mu\nu} + A_5(\Psi) g^{\mu\nu}\hat{\nabla}_\mu\hat{\nabla}_\nu \Psi
+A_6(\Psi)\,.
\ee
For generality, we included also the nonminimal coupling to curvature, $A_1(\Psi)$. The Palatini variation of
this lagrangian was done by Burton and Mann in \cite{Burton:1997pe}. Here we consider the case that the lagrangian is quadratic in the field $\Psi$,
\be
A_1(\Psi)=\frac{1}{2}\lp \frac{1}{\kappa^2}+\al\Psi^2\rp\,,\quad
A_2(\Psi)=\al_2\,,\quad
A_3(\Psi)=\al_3\Psi\,,\quad
A_4(\Psi)=\al_4\Psi\,,\quad
A_5(\Psi)=\al_5\Psi\,,\quad
A_6(\Psi)=-\frac{1}{2}m^2\Psi^2\,.
\ee
By using conformal relations like (\ref{conformal}), (\ref{ricci}) and partial integrations, we find that the
theory can be
equivalently written in the simple form
\be \label{theory2}
2L = \lp \frac{1}{\kappa^2} + \al\Psi^2 \rp \lp R - \kappa^2\lp\partial\varphi\rp^2 \rp
- \beta\lp\partial \Psi\rp^2 + 2\gamma \kappa\Psi\Psi_{,\al}\varphi^{,\al} - m^2\Psi^2\,,
\ee
where $\bt = (\al_5-\al_2)/2$ and
\be
\ga = \frac{1}{2\sqrt{(D-1)(D-2)}}\lb (D-1)\al+2D\al_3+2\al_4+(D-2)\al_5 \rb\,.
\ee
When $\al_5-\al_2>0$, the kinetic term of the
field can be made canonical, $\bt=1$, by rescaling the field. $m$ is the mass.
Thus it is clear that there is essentially just one new parameter, $\ga$, which is due to the presence of
non-metricity. Its effect is to introduce a kinetic mixing between the two scalar degrees of freedom.
Additionally, a nonminimal coupling $\al \neq 0$ results now also in a derivative interaction.

\section{Alternative descriptions}
\label{einstein}

\subsection{Einstein frame}

The rescaling
\be
\tilde{g}_{\mu\nu}=(1+\alpha\kappa^2\Psi^2)^\frac{2}{D-2}g_{\mu\nu}
\ee
brings the lagrangian into the form
\be
\tilde{L}=\frac{1}{2\kappa^2}\tilde{R}+\tilde{X}-\tilde{V}\,,
\ee
where the components of the kinetic matrix defined by $\tilde{X}=-\frac{1}{2}G_{IJ}\tilde{g}^{\mu\nu}\phi^I_{,\mu}\phi^J_{,\nu}$ are given by
\be
G_{\varphi\varphi}=1\,,\quad G_{\varphi\Psi}=G_{\Psi\varphi}=-\frac{\ga\kappa\Psi}{1+\al\kappa^2\Psi^2}\,,\quad G_{\Psi\Psi}=\frac{1}{1+\al\kappa^2\Psi^2}\lb\beta+4\frac{D-1}{D-2}\al\kappa^2\Psi^2\rb\,.
\ee
The potential is
\be
\tilde{V}(\Psi)=(1+\alpha\kappa^2\Psi^2)^\frac{2}{D-2}V(\Psi)\,.
\ee
In this paper we however focus on the case $\al=0$ where the frames coincide.

\subsection{Diagonal frame}
 \label{diagonalization}

To clarify the field content of our the case (\ref{theory}), let us introduce the linear combinations
\be
\phi_\pm=\Psi\pm\varphi\,.
\ee
The Lagrangian can then be written, in terms of the eigenfunctions
\be \label{eigen}
\la_\pm 
= \frac{1}{2}\pm\frac{\ga\ka}{4}\lp\phi_++\phi_-\rp\,,
\ee
as
\be
2L=R-\la_+\lp\partial\phi_-\rp^2 - \la_-\lp\partial\phi_+\rp^2 - 2V\lp\frac{\phi_+ + \phi_-}{2}\rp\,.
\ee
So the diagonalized fields have a nonminimal interaction and both are nontrivially coupled kinetically.
Below ''the barrier'' $\gamma\kappa\Psi<1$ there are no ghosts.

\bibliography{refs}

\end{document}